\documentclass[prb,twocolumn,showpacs,floatfix]{revtex4}

\usepackage{graphicx}
\usepackage{dcolumn}
\usepackage{bm}

\begin{document}
\title{Universality of \textit{ac}-conduction in anisotropic disordered
 systems: An effective medium approximation study}
\author{Sebastian Bustingorry}
\affiliation{Centro At\'omico Bariloche, 8400 Bariloche, R\'{\i}o
Negro, Argentina}

\date{\today}

\begin{abstract}
Anisotropic disordered systems are studied in this work within the random 
barrier model. In such systems the transition probabilities in
different directions have different probability density
functions. The frequency-dependent conductivity at low temperatures is obtained
using an effective medium approximation.
It is shown that the isotropic universal \textit{ac}-conduction
law, $\widetilde{u}=\widetilde{\sigma} \ln \widetilde{\sigma}$, is
recovered if properly scaled conductivity ($\widetilde{\sigma}$) and frequency 
($\widetilde{u}$) variables are used.
\end{abstract}


\maketitle

\section{Introduction}

In the past years, \textit{ac}-conduction in isotropic disordered
systems has been extensively studied.
\cite{fish86,dyr93b,dyr93,dyre94,schr00,dyreR,schrT,roli97,side99,ghos99,roli00b}
Several experiments support the existence of an
universal function governing the conductivity-frequency relation
in a variety of materials, in either electronic
\cite{schr00,dyreR,schrT} or ionic
\cite{schr00,dyreR,roli97,side99,ghos99,roli00b} systems. This
universality is also supported by several theoretical
studies.\cite{dyreR} One of the simplest models for studying conduction in
disordered systems is the random barrier model, in which the
energy barriers joining sites of a given network are selected at
random from a given probability density function (PDF). For this
model the universality of \textit{ac}-conduction at low
temperatures is well established.
Dyre and coworkers studied the isotropic random barrier model 
within the effective medium approximation (EMA).
\cite{dyre94,dyreR,schr00,dyre95}
They found that an universal function is arrived at in a
low-temperature--small-frequency expansion; which corresponds to the
solution of the equation
\begin{equation}
\label{unilaw}
\widetilde{u}=\widetilde{\sigma} \ln
\widetilde{\sigma},
\end{equation}
where  $\widetilde{\sigma}=\sigma(u)/\sigma(0)$ and $\widetilde{u}=u \widetilde{\beta} \ln
\widetilde{\beta}/\sigma(0)$ are scaled conductivity and frequency 
variables respectively (with $\widetilde{\beta}\propto \beta=1/k_BT$, see below).
This equation is universal in the sense that it does not depend on
the characteristic disorder of the system. 
The same equation had been previously arrived at with other approximations,
\textit{e.g.} the macroscopic\cite{fish86,dyr93b,dyr93} and hopping
models.\cite{bryk80}
In addition, other approaches such as the percolation path approximation, the
diffusion cluster approximation and the velocity auto-correlation method, 
predict universal functions which present a better data collapse.
\cite{dyreR,schr00,schrT} 
However, EMA offers a simple systematic tool to study 
\textit{ac}-conductivity of disordered systems analytically which, in addition,
gives the expected universal behavior.

The conductivity properties of anisotropic disordered systems have also
attracted attention in the last years.
\cite{parr87,tole92,gall94,reye00,cacerey,saad02,huang02,bust00,
bust02,bust03,bust04}
Two relevant examples of anisotropic disordered systems are the
superconductor cuprates and porous reservoir rocks. In the first case, 
conductivity properties are strongly anisotropic, with a remarkable
difference between the conductivity in the $ab$ plane and along the
$c$ axis.
In the second example, a relation between permeability and electrical 
conductivity in isotropic fluid-saturated porous media is well established,
\cite{john86,john87,avel91} and a universal behavior for the dynamical 
permeability, analog to Eq. (\ref{unilaw}), was observed
numerical and experimentally. \cite{shen88,char88}
Since anisotropy is a key characteristic of porous media and 
fractured rock,\cite{sahimiL}
the behavior of anisotropic frequency-dependent conductivity in disordered 
media and its relation with the permeability tensor \cite{avel91} is of key
interest.

In view of these and others examples of anisotropic disordered media, is
necessary to find a relation between anisotropic conductivity and frequency.
The approach used
here is to use the ideas and concepts used in isotropic problems and apply them
to anisotropic systems. In this sense, the main purpose of this work is to
extend and generalize the universal law, Eq. (\ref{unilaw}), to anisotropic
systems. An anisotropic random barrier model in two-dimensions is 
studied within an effective medium theory. 
The analytical procedure used by Dyre \cite{dyre94} is closely followed in order
to obtain a consistent low-temperature--small-frequency expansion for the
anisotropic \textit{ac}-conductivity.
Here, from the analysis of the low temperature limit, the isotropic universal 
function is recovered for properly scaled variables relating the 
conductivity in both directions.
The paper is organized as follows. In Sec. \ref{s:model} the anisotropic random
barrier model is described and previous zero-frequency results are summarized.
Section \ref{s:ema} briefly presents the main
features of the anisotropic EMA and in Sec. \ref{s:univ} the 
low-temperature--small-frequency expansion is performed. 
Finally, a summary is presented in Sec. \ref{s:summary}.

\section{Background}\label{s:model}

In the anisotropic random barrier model considered here, equal
energy minima form a square lattice with its four
nearest-neighbors separated by energy barriers, whose heights are
randomly selected from anisotropic PDFs. Let $1$ and $2$ be the
main directions of the square lattice, and the PDFs in each
direction $\rho_1(E_1)$ and $\rho_2(E_2)$. Once the energy barrier
$E_{\alpha \beta}$ between two nearest-neighbor sites $\alpha$ and $\beta$ 
is selected from the corresponding PDF, the transition
rate between sites $\alpha$ and $\beta$ is given by
$\omega_{\alpha \beta}=\omega _{0}exp(-\beta E_{\alpha \beta})$, 
where $\omega_0$ is the constant jump rate and $\beta=1/k_BT$
is the inverse temperature, with $k_B$ being the Boltzmann
constant.
Since $E_{\alpha \beta}=E_{\beta \alpha}$, the forward 
($\alpha \rightarrow \beta$) and backward ($\beta \rightarrow \alpha$)
jumps have the same transition rate.
The energy PDFs, $\rho_1(E_1)$ and $\rho_2(E_2)$, are related to the
transition rate PDFs, $\nu_1(\omega_1)$ and $\nu_2(\omega_2)$, through the
corresponding transformation of random variables.

The zero-frequency conductivity of the anisotropic random barrier
model in two dimensions was recently studied by using an
anisotropic generalization of the EMA.\cite{bust04} 
The low temperature conductivity in each
direction was shown to follow Arrhenius laws with the same
activation energy $E_c$, which is determined by the anisotropic
percolation properties of the lattice. For the square lattice
studied, the bond percolation threshold is the critical surface
$p_1+p_2-1=0$,\cite{grimmet} where $p_i$ represents the probability of
having a conducting link between two nearest-neighbor sites in
the $i$ direction. This implies an activation energy given by
\cite{bust04}
\begin{equation}
\label{ecani} \int_0^{E_c}{\rho_1(E_1)dE_1} +
\int_0^{E_c}{\rho_2(E_2)dE_2} = 1.
\end{equation}
The zero-frequency conductivities, $\sigma_i(u=0)$, in each direction are thus 
given by
\begin{eqnarray}
\label{sigma0}
\sigma_1(0)&=&\gamma_{12}\omega_0 a^2 e^{-\beta E_c},\nonumber \\*
\sigma_2(0)&=&\gamma_{21}\omega_0 a^2 e^{-\beta E_c},
\end{eqnarray}
where $a$ is the lattice constant and the prefactor 
\begin{equation}
\label{gamma}
\gamma_{12}=\gamma_{21}^{-1}=\frac{\int\limits_{0}^{E_c}\rho_1(E_1) dE_1}
{\int\limits_{0}^{E_c}\rho_2(E_2) dE_2}.
\end{equation}
Note that, at low temperatures, the anisotropic character of the system 
reflects only in the prefactors $\gamma_{12}$ and $\gamma_{21}$ of the 
zero-frequency conductivity.

\section{Anisotropic EMA} \label{s:ema}

The EMA consists in averaging the effects of disorder by defining
an effective medium with effective transition rates, which depend
on the Laplace variable $u$. These effective transition rates are
self-consistently determined by the requirement that the
difference between the propagator of the impurity and homogeneous
problems should average to zero.
\cite{brug35,brug35b,brug36,brug39,land77,alexR,odag81,webm81,summ81}
In anisotropic problems, two effective transition rates, 
one for each direction, are introduced.
The effective frequency-dependent conductivities of the disordered
system, $\sigma_1(u)$ and $\sigma_2(u)$, are proportional to
the effective transition rates. \cite{parr87,reye00}
A \textit{rationalized} unit system is used where all the prefactors are 
absorbed in the definition of effective conductivities, so that they are 
equivalent to effective transition rates.
\cite{dyre94}
These effective conductivities
are then determined by two self-consistent conditions:
\cite{parr87,reye00}
\begin{eqnarray}
\label{aveani1}
\left\langle {\frac {\sigma_1-\omega_1}
{1+2\left( \sigma_1-\omega_1\right)\left[ G^{1} (u)-G^{0} (u)\right]}}
\right\rangle _{\nu_1 (\omega_1)}=0, \nonumber \\*
\left\langle {\frac {\sigma_2-\omega_2}
{1+2\left( \sigma_2-\omega_2\right)\left[ G^{2} (u)-G^{0} (u)\right] }}
\right\rangle _{\nu_2 (\omega_2)}=0.
\end{eqnarray}
Here, $G^{1(2)}$ and $G^{0}$ represent the non-perturbed anisotropic Greens
functions related to the probabilities of moving from the origin to one of its
nearest neighbors in the $1(2)$ direction and the return probability,
respectively.
The impure bond connects two nearest neighbor sites of the lattice whose
transition rates are equal to $\omega_{1}$ if the impure bond lies in the $1$
direction and $\omega_{2}$ if the impure bond is in the other direction.
The angular brackets denote averaging over the corresponding 
transition rate PDFs.

The real and imaginary parts of the frequency-dependent conductivity are obtained
considering that the Laplace frequency $u$ is actually an imaginary frequency
related to the real frequency by $u=i s$.
\cite{dyre94,odag81}
However, the Laplace frequency picture has proved to be a useful
simplification for studying frequency-dependent conductivity.
\cite{dyre94,dyreR}

\section{Universal frequency-dependent conductivity} \label{s:univ}

In this section a low-temperature--small-frequency expansion \cite{dyre94} 
of the set of Eqs. (\ref{aveani1}) is performed. 
In the following, the subscripts $i,j=1,2$ are used to represent the two different
directions of the lattice, noting that one of the self-consistent conditions in
Eqs. (\ref{aveani1}) is obtained interchanging the subscripts $i$ and $j$ in the
other equation.
To the lowest order in $u$, the difference between the Greens
functions in the $i$ direction appearing in the denominator of 
Eqs. (\ref{aveani1}) may be written as \cite{reye00}
\begin{equation}
\left( G^{0}-G^{i} \right)_{u \rightarrow 0}= \frac{f_{ij}}{2
\sigma_i}+ \frac{u g_{ij}}{2 \sigma_i},
\end{equation}
with
\begin{equation}
f_{ij}=\frac{2}{\pi}\ \arctan \sqrt{\frac{\sigma _{i}}{\sigma _{j}}},
\end{equation}
and
\begin{equation}
\label{gij}
g_{ij}=\frac{1}{4 \pi \sqrt{\sigma_i \sigma_j}} \ln{\frac{64
\sigma_i \sigma_j}{u\left(\sigma_i+\sigma_j\right)}}.
\end{equation}
Using this expansion for the Greens functions near $u=0$, the
frequency-dependent self-consistent condition Eqs. (\ref{aveani1})
read
\begin{equation}
\label{equ0} 
\left\langle \frac{\omega_i-\sigma_i}{\omega_i
+\left[ \left( f_{ij}-u g_{ij} \right)^{-1} -1\right] \sigma _{i}}
\right\rangle _{\nu _{i}\left(\omega_{i}\right)}=0,
\end{equation}
In order to emphasize the energy dependence of the transition rates, one may
average over the energy distributions and write, rearranging terms in the previous
equation,
\begin{equation}
\label{equ1} 
\left\langle \frac{1}{\omega_i + \left[\left(f_{ij}-u
g_{ij} \right)^{-1} -1\right] \sigma _{i}} \right\rangle
_{\rho_{i}\left(E_{i}\right)}= \frac{f_{ij}-u g_{ij}}{\sigma_i}.
\end{equation}

Two regimes may be identified for each direction:
$\omega_i\left(E_i\right)\ll\left[\left(f_{ij}-u g_{ij}
\right)^{-1} -1\right] \sigma _{i}$, and 
$\omega_i\left(E_i\right)\gg\left[\left(f_{ij}-u g_{ij}
\right)^{-1} -1\right] \sigma _{i}$. \cite{bust04}
The energy separating these two cases,
$E^g_i(u)$, is defined by
\begin{equation}
\omega_i\left[E^g_i(u)\right]=\left[\left(f_{ij}-u g_{ij}
\right)^{-1} -1\right] \sigma _{i},
\end{equation}
which yields
\begin{equation}
\label{egdeu}
E^g_i\left(u\right)=-\frac{1}{\beta} \ln\{\left[
\left( f_{ij}-u g_{ij} \right)^{-1}-1
\right]\frac{\sigma_i}{\omega_0}\}.
\end{equation}
Using these different ranges of energy, separated by $E^g_i$, and the fact that
transition rates and energies are related through an Arrhenius law, the
integral of the energy average in Eq. (\ref{equ1}) may be approximated,
and the self-consistent condition, Eq. (\ref{equ1}), then reads
\begin{equation}
\int\limits_{E^g_i(u)}^{\infty}
\frac{\rho_i(E_i)}{\left[\left(f_{ij}-u
g_{ij}\right)^{-1}-1\right] \sigma_i}dE_i=\frac{f_{ij}-u
g_{ij}}{\sigma_i},
\end{equation}
or, equivalently,
\begin{equation}
\label{equ2} 
\int\limits_{E^g_i(u)}^{\infty}
\rho_i(E_i)dE_i=1-f_{ij}+u g_{ij}.
\end{equation}

In the zero-frequency case, $E^g_i$ becomes the same for the two
directions,\cite{bust04} and is given by Eq. (\ref{ecani}) with
$E_i^g(u=0)\equiv E_c$. Noting that for $u=0$, $f_{ij}$ is given by
\begin{equation}
\label{fij0}
f_{ij}=\int\limits_{0}^{E_c} \rho_i(E_i)dE_i.
\end{equation}
and inserting Eq. (\ref{fij0}) into  Eq. (\ref{equ2}) one obtains
\begin{equation}
\label{equ3a}
\int\limits_{E^g_i(u)}^{E_c} \rho_i(E_i)dE_i=u g_{ij},
\end{equation}
which may be approximated by
\begin{equation}
\label{equ3b}
\int\limits_{E^g_i(u)}^{E_c} \rho_i(E_i)dE_i \simeq q_i
\left[E^g_i(0)-E^g_i(u)\right],
\end{equation}
with
$q_i=\rho_i\left[E^g_i(0)\right]=\rho_i\left[E_c\right]$.
Evaluating $E^g_i(u)$ from Eq. (\ref{egdeu}) and combining
Eq. (\ref{equ3a}) with Eq. (\ref{equ3b}) one obtains, 
to the lowest order in $u$,
\begin{equation}
u g_{ij}=\frac{q_i}{\beta} \ln \frac{\left[ \left(f_{ij}-u
g_{ij} \right)^{-1} -1 \right] \sigma_i}{\left(f_{ij}^{-1}-1
\right) \sigma_i(0)} \simeq \frac{q_i}{\beta} \ln
\frac{\sigma_i}{\sigma_i(0)},
\end{equation}
where $\sigma_i(0)$ is the zero-frequency conductivity in the $i$
direction, and is given by Eq. (\ref{sigma0}). 
Then, using the definition of $g_{ij}$, Eq. (\ref{gij}), the previous equation
my be written as
\begin{equation}
\label{sideu}
\ln \frac{\sigma_i}{\sigma_i(0)}=\widetilde{\beta_i}
\frac{u}{\sqrt{\sigma_i \sigma_j}} \ln \frac{64 \sigma_i
\sigma_j}{u(\sigma_i + \sigma_j)},
\end{equation}
with $\widetilde{\beta_i}=\beta/(4 \pi q_i)$.
This last equation gives the general frequency-dependent
conductivity for all temperatures and in the small frequency limit.

In order to obtain the low temperature limit for the frequency-dependent
conductivity, scaled conductivity and frequency variables for
each direction are introduced, namely
\begin{equation}
\widetilde{\sigma_i}=\frac{\sigma_i}{\sqrt{\sigma_i(0)
\sigma_j(0)}},
\end{equation}
and
\begin{equation}
\widetilde{u_i}=\frac{\widetilde{\beta_i} \ln \widetilde{\beta_i}}{\sqrt{\sigma_i(0)
\sigma_j(0)}} u.
\end{equation}
With this scaled variables Eq. (\ref{sideu}) may be written as
\begin{eqnarray}
\ln \widetilde{\sigma_i} + \frac{1}{2}\ln \frac{\sigma_j(0)}{\sigma_i(0)}=
\nonumber \\
\frac{\widetilde{u_i}}{\sqrt{\widetilde{\sigma_i}\widetilde{\sigma_j}}
\ln \widetilde{\beta_i}} 
\left[\ln \frac{64\widetilde{\sigma_i}\widetilde{\sigma_j}}
{\widetilde{u_i}(\widetilde{\sigma_i}+\widetilde{\sigma_j})}
+\ln \widetilde{\beta_i}+ \ln \left(\ln \widetilde{\beta_i}\right)
\right].
\end{eqnarray}
By taking the low temperature limit $\beta \rightarrow \infty$ for fixed
$\widetilde{u_i}$ and $\widetilde{\sigma_i}$,
the following set  of coupled equations is arrived at for the scaled
conductivities as functions of the scaled frequencies:
\begin{eqnarray}
\label{unilawani} 
\widetilde{u_1}&=&\sqrt{\widetilde{\sigma_1}
\widetilde{\sigma_2}} (\ln \widetilde{\sigma_1}+ \ln \gamma_{21})
\nonumber \\* 
\widetilde{u_2}&=&\sqrt{\widetilde{\sigma_2}
\widetilde{\sigma_1}} (\ln \widetilde{\sigma_2}+ \ln \gamma_{12}),
\end{eqnarray}
where Eqs. (\ref{sigma0}) and (\ref{gamma}) were used.

This set of equations represent the complex relation between the
conductivity in each direction. Although they may be regarded as
universal equations for the scaled conductivities, the use of two
different frequencies, one for each direction, is not suitable for a 
frequency-dependent description. 
In addition, the terms containing $\gamma_{12}$ still depend on the specific 
PDFs used.
Still, Eqs. (\ref{unilawani}) are useful to obtain the
frequency-dependent conductivities: given the PDFs for each direction, $E_c$ and
$\gamma_{12}$ are calculated through Eqs. (\ref{ecani}) and (\ref{gamma}), then
through the definitions of $\widetilde{u_i}$ and $\widetilde{\beta_i}$ the 
scaled conductivities in each direction may be calculated numerically by solving
the set of coupled Eqs. (\ref{unilawani}).

As the analytical derivation of Eqs. (\ref{unilawani}) closely follows the
previous derivation of Eq. (\ref{unilaw}) from EMA, \cite{dyre94} the isotropic
result is obviously recovered by setting the same PDF, $\rho(E)$, for the two
directions, which implies 
$\widetilde{\sigma_1}=\widetilde{\sigma_2}=\widetilde{\sigma}=\sigma(u)/\sigma(0)$
and $\widetilde{u_1}=\widetilde{u_2}=\widetilde{u}=u \widetilde{\beta} \ln
\widetilde{\beta}/\sigma(0)$, with $\widetilde{\beta}=\beta/[4\pi\rho(E_c)]$.
However, by adding the two Eqs. (\ref{unilawani}) 
and setting $\widetilde{u}=(\widetilde{u_1}+\widetilde{u_2})/2$ and
$\widetilde{\sigma}=\sqrt{\widetilde{\sigma_1}\widetilde{\sigma_2}}$, 
expression (\ref{unilaw}) is obtained again.
This is a non-trivial result and establish that although the anisotropic
conductivities are coupled through the complex relation given by 
Eqs. (\ref{unilawani}), the frequency dependence of the geometric mean
conductivity, $\widetilde{\sigma}$, in the anisotropic problem is simple given
by Eq. (\ref{unilaw}). This result given by the EMA should
be tested by more complex theories.

\section{Summary} \label{s:summary}

In summary, by means of the frequency-dependent EMA,
conductivity in an anisotropic random barrier model has been studied. 
It was shown that in a small-frequency expansion, the low temperature 
limit is characterized by an universal law, \textit{i.e.} independent of the
anisotropic PDFs, relating scaled conductivity and frequency variables. 
This scaled quantities are obtained by a proper combination of the
conductivities and energy properties in each direction. 
Although the universal law is obtained for a two-dimensional system,
it is expected to hold in three-dimensions for appropriately scaled variables.
Direct comparison with experiments on the conductivity of superconductor
cuprates is not possible at present because the available data corresponds,
to the best of the author's knowledge, to conductivity measurements in only 
one of the anisotropic directions of the system. Unfortunately, for a comparison
with the scaling function, the conductivity in both directions should be
available.
The results presented can be of relevance for a complete theory of the
anisotropic permeability of porous reservoir rocks. Given the relation between
the anisotropic conductivity and permeability tensors, \cite{avel91} the
dynamical permeability in each relevant direction can be obtained through Eqs.
(\ref{unilawani}).
Finally, is worth noting that a perfect agreement between the EMA universal law
and experimental or simulation data should not be expected, as this is the case
even for isotropic problems.\cite{dyreR,schr00} 
Other theoretical and
simulation methods were shown to predict a better universal law,
which collapses experimental data from various disordered
systems.\cite{dyreR,schr00} 
However, EMA still provides a simple
analytical tool for a first exploration of the properties of
\textit{ac}-conduction. In the present work, a first insight
on the emergence of an universal law for anisotropic disorder
systems has been presented.

\section*{Acknowledgments}

This work was financially supported by \mbox{CONICET}, Argentina.

\bibliography{barbm}

\end{document}